\documentclass[a4paper,12pt]{article}
\pdfoutput=1
\usepackage[round, comma]{natbib} 
\usepackage[utf8]{inputenc}
\usepackage[T1]{fontenc} 
 
\usepackage[nooneline,bf]{subfigure}  
\usepackage[hyphens]{url}
\usepackage{hyperref}  
\usepackage{times}
\usepackage{graphicx}
\usepackage{wordlike}
\begin{document}

\title{Inconsistencies between long-term trends in storminess derived
from the 20CR reanalysis and observations}
\author{O. Krueger\thanks{E-mail: oliver.krueger@hzg.de}, F. Schenk, F. Feser, and R. Weisse}
\date{\small{Institute for Coastal Research, Helmholtz-Zentrum Geesthacht, Max-Planck-Str. 1, 21502 Geesthacht, Germany}}

\maketitle

\begin{abstract}
Global atmospheric reanalyses have become a common tool for both the validation of climate models and diagnostic studies, such as assessing climate variability and long-term trends. Presently, the 20th Century Reanalysis (20CR), which assimilates only surface pressure reports, sea-ice, and sea surface temperature distributions, represents the longest global reanalysis dataset available covering the period from 1871 to the present. Currently, the 20CR dataset is extensively used for the assessment of climate variability and trends. Here, we compare the variability and long-term trends in Northeast Atlantic storminess derived from 20CR and from observations. A well established storm index derived from pressure observations over a relatively densely monitored marine area is used. It is found that both, variability and long-term trends derived from 20CR and from observations, are inconsistent. In particular, both time series show opposing trends during the first half of the 20th century. Only for the more recent periods both storm indices share a similar behavior. While the variability and long-term trend derived from the observations are supported by a number of independent data and analyses, the behavior shown by 20CR is quite different, indicating substantial inhomogeneities in the reanalysis most likely caused by the increasing number of observations assimilated into 20CR over time. The latter makes 20CR likely unsuitable for the identification of trends in storminess in the earlier part of the record at least over the Northeast Atlantic. Our results imply and reconfirm previous findings that care is needed in general, when global reanalyses are used to assess long-term changes.
\end{abstract}


\noindent\textit{This document has been published in Journal of Climate and can be found online at \url{http://dx.doi.org/10.1175/JCLI-D-12-00309.1}.}

\end{document}